\documentstyle{mn}
\input psfig.sty
\input epsf.sty
\begin{document}
\title{Magnetically collimated jets with high mass flux}
\author[S. Bogovalov, K. Tsinganos]{S. Bogovalov$^1$ and K. Tsinganos$^2$
\\$^1$Astrophysics Institute at the 
Moscow State Engineering Physics Institute, Moscow 11549, Russia\\
$^2$Department of Physics, University of Crete,  
710 03 Heraklion, Crete, Greece} 
\date{February 14, 2001} 
\maketitle 
\begin{abstract} 
Recent numerical simulations and analytical models of magnetically 
collimated plasma outflows from a uniformly rotating central 
gravitating object and/or a Keplerian accretion disk have shown that 
relatively low mass and magnetic fluxes reside in the produced jet. 
Observations however indicate that in some cases, as in jets of 
YSO's, the collimated outflow carries higher fluxes than these 
simulations predict.   
A solution to this problem is proposed here by assuming that jets 
with high mass flux originate in 
a central source which produces a noncollimated outflow provided that 
this source is surrounded by a rapidly rotating accretion disk. 
The relatively faster rotating  disk produces  a collimated 
wind which then forces all the enclosed outflow from the central
source to be collimated too. 
This conclusion is confirmed by self-consistent numerical solutions 
of the full set of the MHD equations. 
\end{abstract} 
\section{ Introduction}

There has been already accumulated enough evidence for a {\it correlation} 
bewtween  
the presence of collimated astrophysical outflows and accretion disks, not only 
in star formation regions (for a review see e.g., K\"onigl \& Pudritz  
2000), but also in other galactic or extragalactic jet-producing  
sources (for a review see e.g., Livio 1999).  
For example, in star forming regions, an apparent correlation is found between 
accretion diagnostics and outflow signatures
(Hartigan et al. 1995).
Hence, in such cases astrophysical jets are believed to be fed by the material 
of an accretion disk which surrounds the central object. 
Most jets are also believed to be powered by the gravitational energy which 
is released in the accretion process. 
For example, in the case of jets associated with YSO's 
it is well known that the observed mechanical luminosity in the bipolar outflows 
is typically a factor $\sim 10^2$ higher than the total radiant luminosity of 
the embedded central object (Lada 1985), a fact that seems to rule out  
radiative acceleration of those jets. Furthermore, the kinetic luminosity of the 
outflow ($ \sim \dot M_{\rm wind} v_w^2$) seems to be a fraction $\sim 0.1$ 
of the rate at which energy is released by accretion ($\dot M_{\rm acc} v_K^2$), 
if $\dot M_{\rm acc} \sim 10 \dot M_{\rm wind}$ and 
the outflow speed is of the order of the Kepler speed $v_K$.  Such a high ejection 
efficiency is most naturally understood if the jets are driven magnetically  
(K\"onigl \& Pudritz 2000).  

Magnetic fields seem to be also implicated to the most striking feature of jets, 
e.g., their indeed high degree of collimation. The most dramatic manifestation 
of such collimation is the HST observed disk/jet system HH30 where the jet appears 
to be collimated within a cone of opening angle 3$^o$ and can be traced to within 30 
AU from the star (Burrows et al. 1996).
From the theoretical point of view, it has been already demonstrated that the 
magnetic hoop stress can naturally collimate a plasma outflow
(Heyvaerts \& Norman 1989, Bogovalov 1995). 
In steady-state and axisymmetric analytical models  
exact solutions of the full set of the MHD equations have been obtained where 
a wind-type outflow starting either radially from a spherical source or 
non vertically from a disk 
becomes eventually cylindrically collimated after crossing the fast critical surface,
provided that the source of the outflow is an efficient magnetic rotator 
(Sauty \& Tsinganos 1994, Vlahakis \& Tsinganos 1998,  Sauty et al. 1999). 
On the other hand, in time-dependent simulations a similar result has been 
obtained. For instance, in Bogovalov \& Tsinganos (1999, 
henceforth Paper I), 
when an initially nonrotating and radial magnetosphere with a uniform radial  
plasma outflow starts rotating, significant flow collimation is obtained {\it if} the 
corotating speed at the Alfv\'en distance is larger than the initial 
flow speed.            
   
A serious limitation however of the previous simulations of magnetic collimation 
is that only a tiny fraction of order $\sim 1\%$ of the mass and magnetic flux 
of the originally radial wind ends up collimated inside the jet (Paper I). 
Similarly, in analytical models if the source of the wind is a stellar 
surface and the disk does not feed the outflow with mass and magnetic flux, 
very low wind- and jet-mass loss rates ($\dot M_{wind}$, $\dot M_{j}$) 
are obtained. 
However, in outflows associated with YSO current estimates place
$\dot M_{\rm jet}$ in the limits 
$\dot M_{\rm jet} \sim 10^{-6} - 10^{-8} M_{\odot}/yr$ (Ray 1996).   
And, the inferred from observations mass loss rates 
of bipolar outflows indicate wind mass loss rates which are also in the range   
of $\dot M_{\rm wind} \sim 10^{-6} - 10^{-8} M_{\odot}/yr$, depending largely 
on the luminosity of the YSO's. Therefore, the mass loss rate in the jet has to 
be a rather large fraction of the mass loss rate in the surrounding wind.

In the present paper we shall show via simulation examples that it is possible 
to have a large fraction of the wind mass loss rate inside the jet, 
for suitable distributions of the angular rotation frequency of the system.  
In the next section 2 we give a detailed comparison of theory and observations 
regarding fluxes in jets while in section 3 we describe and qualitatively sketch our model 
and the numerical method to calculate the time-dependent evolution; then,  
our results are presented for various laws of the distribution 
of the angular rotation frequency with the magnetic flux function. 
A summary and physical discussion of these results is given in 
the last Section 4.

\section{Mass flux in  jets: theory vs observations}
In this section it is shown how the results of numerical simulations of 
collimated MHD outflows are inadequate to reproduce the mass flux which 
observations of some jets seem to indicate.    
\subsection{Theoretical results} 
Magnetized winds posses the important property of {\it selfcollimation}. 
According to a general analysis, as in Heyverts \& Norman (1989), Li et al. (1992),      
Bogovalov (1995) and analytical examples  as in Vlahakis \& Tsinganos (1998), 
at large distances from a rotating 
central object, the flow of a magnetized wind will be partially collimated 
into a jet directed along the axis of rotation, {\it if} there exists a poloidal 
fieldline 
which encloses a finite poloidal current. Hence, it is argued  
that the observed astrophysical jets are collimated by  magnetic stresses. 
Numerical simulations seem to confirm this conclusion. 
For instance, in simulations performed by the method of relaxation as in  
Oyued \& Pudritz (1997), Ustyugova et al. 
(2000) Krasnopolsky et al. (2000), Kudoh et al. (1998) and 
Keppens \& Goedbloed (2000),  steady state solutions have been 
obtained in the nearest zone of the simulation with a dimension comparable to 
the size of the critical surfaces. Nevertheless, although in all those 
studies  collimation 
of the plasma around the axis of rotation has been found, these results
cannot be
directly compared with observations of jets. 

In observations of jets we deal with scales which are rather large compared
with
all scales in the nearest zone, such as, the dimension of the accretion 
disk, the size of the critical surfaces, etc. In the observational scale, the central
source
would appear simply as a point. Therefore, it is important for a direct  
comparison with observations to have a solution of the problem at large 
distances.  
Up to now, the only self-consistent solution of the problem of the MHD wind 
structure at large distances from the central object seems to have been   
obtained in Paper I, where 
it was found in all analysed cases that a narrow jet is indeed 
formed over very large distances. 
However, the mass flux in this jet is only a small fraction of the 
total mass flux of the originally spherical wind.
And, this result appears to be rather general for all cases considered in
the analysis.  In particular, calculations for a wide class of outflows with 
{\it disk-like} laws of rotation and also including winds with a thermal
pressure,
provided a similar result (Tsinganos \& Bogovalov 2000, henceforth Paper II). 

To illustrate that this feature of the simulations is independent of the 
parameters, we  present in Fig. 1 the results of calculations on the jet 
structure in a cold wind from the study of Paper I 
and for an accretion disk-like law of rotation.  
The angular velocity in this case decreases with an increase of the magnetic 
flux, as  
\begin{equation}     
\Omega = \alpha{v_0\over R_a} \exp(-3{\psi\over \psi_{\rm max}}) 
\,,
\end{equation}
where $\psi$ is the magnetic flux through a surface of radius $r$,   
$\psi_{\rm max}$ is the total magnetic flux in the upper hemisphere, 
$v_o$ the constant outflow speed of the initial non rotating star where 
the Alfv\'en spherical distance is $R_a$. 
 
In Fig. 1, the jet structure is shown for two values of the parameter 
$\alpha$, $\alpha=1$ and $\alpha=2$. 
The wind consists of two parts. 
One component is directed along the axis of rotation and represents
the jet with mass flux $\dot M_{\rm jet}$. The second component expands radially
as in ordinary winds. The total outflowing mass flux is
concentrated in the wind. Therefore it is natural to define the 
total mass flux rate as $\dot M_{\rm wind}$. In  all cases  under 
consideration the mass flux is proportional to the magnetic field flux
represented in Fig. 1 by dash-dotted lines. Therefore, the relative mass 
and magnetic field fluxes are the same. Although the radius of the  
jet differs significantly, the relative mass flux appears 
to be the same in the two cases and is of order 1$\%$.  
This result illustrates our general conclusion that for all sets of 
parameters examined, we obtain in the jet a tiny fraction of the total mass 
flux in the wind. The amount of the mass flux in the jet does not depend 
on the distance from the source in the asymptotic regime considered here, 
since the other part of the outflow expands radially. 
\begin{figure}  
\centerline{\psfig{file=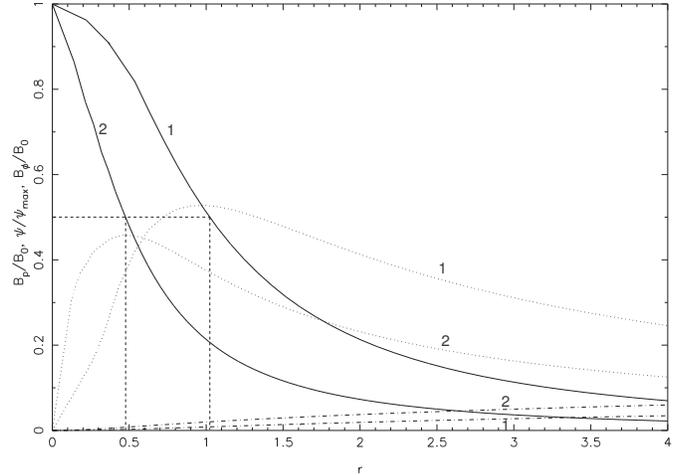,width=9.0truecm,angle=270}} 
\caption{Normalized distributions of poloidal (solid) and  toroidal magnetic 
field (dotted) components together with the relative mass flux (dash-dotted) 
in a simulated jet as a function of the cylindrical distance $r$ from its axis and  
for two values of the parameter $\alpha$ in Eq. (1), $\alpha=1$ in (1) 
and $\alpha=2$ in (2)} 
\end{figure} 

These results are valid for jets spontaneously collimated by magnetic stresses 
from an initially isotropic wind.  
An analysis of the results of simulations of outflows from accretion  
disk-like objects in the nearest zone  
performed by Ustyugova et al. (2000) and Krasnopolsky et al (2000) 
provide evidence that in these cases as well we  have a small fraction 
of the mass flux in the jet in comparison with the total mass flux in the wind. 
For the solution obtained in Ustyugova et al. (2000) this is evident and  
the authors of this paper stress that the collimation of their wind  
is very small, even in the nearest zone.

It is likely that the jet from the wind considered by Krasnopolsky et al. (2000) 
finally will also carry a small relative mass flux because they considered 
a wind from a so-called {\it slow magnetic rotator}. We recall that it has 
been proposed to classify magnetic rotators ejecting a cold plasma into two  
groups, slow and fast magnetic rotators, according to the value of the 
parameter  $\alpha=\Omega R_a/V_0$, where $R_a$ is the initial Alfvenic radius
of the outflow, $V_0$ is the  initial velocity of the plasma and 
$\Omega$ the angular rotation frequency (Paper I). 
Actually, this parameter can be also written as 
$\alpha= 2\pi T_{\rm travel}/T_{\rm rot}$,  
where $T_{\rm travel}$ is the time the plasma spends travelling from the  
base of the outflow to the Alfvenic surface at a radius $R_a$ and $T_{\rm rot}$ 
is the period of rotation. For the slow magnetic rotators, $\alpha < 1$ and the 
plasma leaves the sub Alfvenic region where it can be efficiently accelerated 
and collimated in a time interval which is small compared to the period of 
rotation of the central source.  
That is why winds from slow rotators do not corotate with the ejecting 
object in the subAlfvenic region (no matter if the ejecting object is a star or 
an accretion disk).  
It is exactly this situation that we observe in the solution by Krasnopolsky 
et al. (2000).
It follows from Fig. 3 of this paper that the plasma never corotates 
in their solution. This means that they consider the outflow of plasma 
from a slow rotator with $\alpha < 1$. But the collimation of the ejected plasma  
from slow rotators is not effective and therefore jets  
with a ratio $\dot M_{\rm jet}/\dot M_{\rm wind} \sim 1\%$  
as in Paper I will be produced also.

We conclude that in all available models of magnetized winds, jets with a ratio  
$\dot M_{\rm jet}/\dot M_{\rm wind}$ not higher than $1\%$ are obtained. 
It is interesting to compare this value with the corresponding measured value  
of observed jets from astrophysical objects.  
 
\subsection{Observations}

Although jets have been observed in association with many astrophysical 
objects of different scales and nature (Livio 1999), the most reliable
information on the mass flux in  
jets is obtained apparently only for several YSO's (Hartigan et al. 1995). 
However, even in these objects estimates  of the
ratio $\dot M_{\rm jet}/\dot M_{\rm wind}$  do not seem to be available 
in the literature. There only exist estimates of the ratio
$\dot M_{\rm jet}/\dot M_{\rm accr}$ where $\dot M_{\rm accr}$ is the accretion rate, 
for a few YSO's (Hartigan et al. 1995).  
Early estimates of $\dot M_{j}$ were too low, as they underestimated the 
neutral fraction in the jet (Raga et al. 1990). More recent calculations 
suggest that more than 90$\%$ of the jet may be neutral (Hartigan et al. 1994, 
Morse et al. 1995).   
The luminosity of jets in some forbidden lines, basically O I, indicates 
mass-loss rates in the interval $10^{-8} - 10^{-10} M_{\odot} yr^{-1}$ while 
the accretion mass rate 
derived from the infrared luminosity of the disks is in the range of 
$10^{-6} - 10^{-8} M_{\odot} yr^{-1}$. 
The largest value of the ratio $\dot M_{\rm jet}/\dot M_{\rm accr}$ is about $7\%$
for HH 47 (Hartigan et al. 1994) which demonstrates that at least 
some YSO's have a ratio $\dot M_{\rm jet}/\dot M_{\rm wind} >1\% $ since for YSO's 
$\dot M_{\rm wind} < \dot M_{\rm accr}$ (Pelletier \& Pudritz 1992,  
Ferreira \& Pelletier 1995). 

Thus, from the above considerations it follows that there seem to be at least 
some cases of jets where the mass flux in the jet is comparable to the mass 
flux in the surrounding wind.  
As a result, the theory of spontaneous magnetic collimation of jets should be 
able to incorporate the possibility to form jets with such high ratio of mass 
loss in the jet to the mass loss in the wind.

\section{Jets with high mass flux}
In this section we substantiate our model with results of numerical simulations 
for outflows from a central source collimated by an envelope of rapidly rotating 
disk-wind.   

\subsection{Qualitative analysis}

The rather dramatic discrepancy of the prediction of the theory of magnetic 
collimation with the corresponding observations on the ratio of the mass flux 
in the jet and the wind raises a question on the ability 
of the mechanism of magnetic collimation to provide the collimation of high 
mass fluxes into the jets. 
Qualitatevely the answer to this question follows from an asymptotic theorem  
formulated in Heyvaerts \& Norman (1989), Li et al. (1992) and 
Bogovalov (1995). 
According to this theorem, asymptotically, i.e., at large distances from a 
rotating central source emitting a magnetized outflow, there exists 
at least one streamline of the wind with $\Omega(\psi) \ne 0$ which is 
directed exactly along the axis of rotation.

\begin{figure} 
\centerline{\hspace{1.0cm}\psfig{file=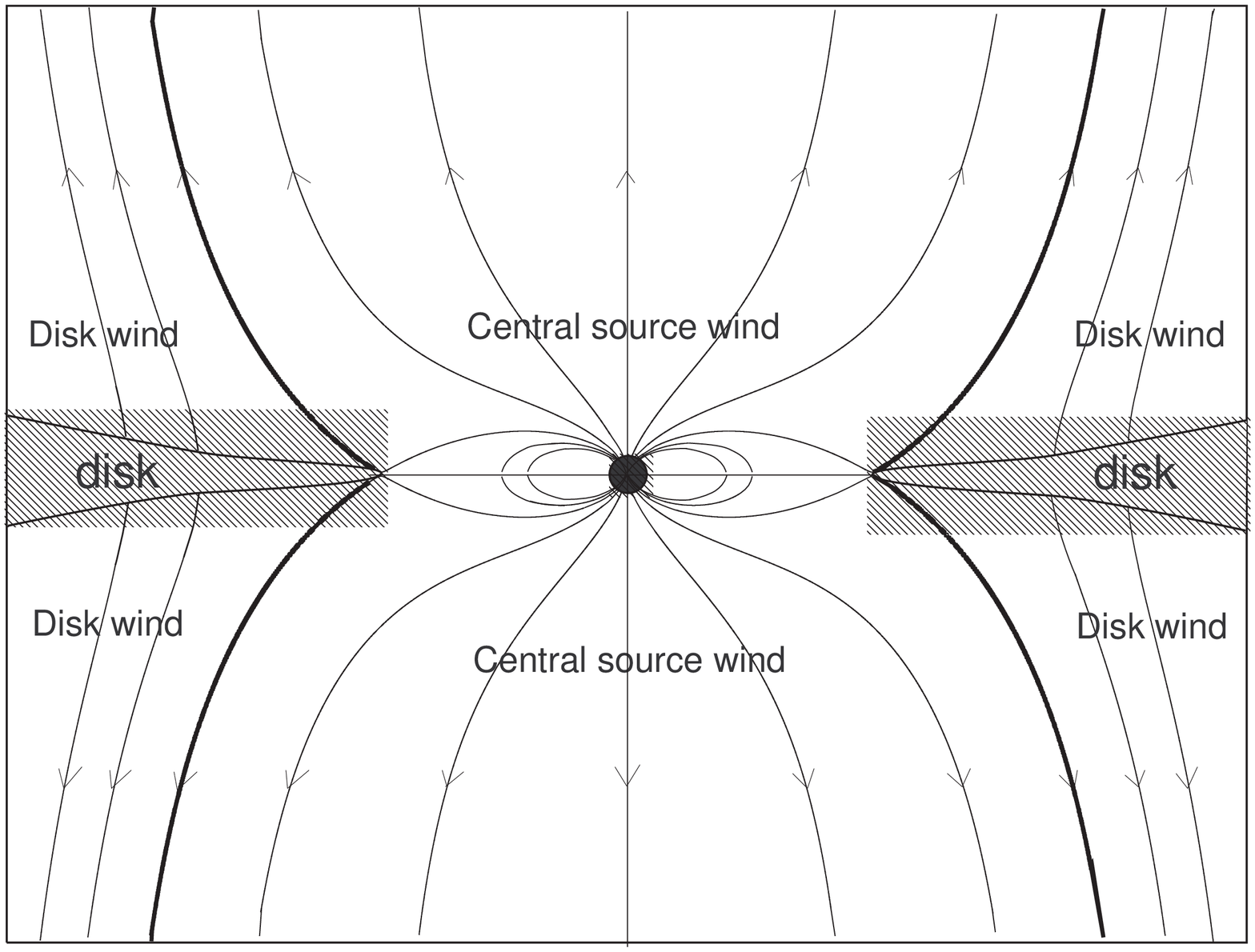,width=9.0truecm,angle=270}}
\caption{Illustration of the model under consideration. A slowly
rotating central source emits a roughly radially expanding at the base outflow  
which is forced to collimate by the wind from the fast rotating 
inner edges of the surrounding accretion disk.}
\end{figure}

Let us consider the following simple and limiting case. Assume that the source of the 
jet consists of a Keplerian disk 
which launches a magnetized wind and a central source emitting from its 
base an isotropic wind (Fig 2.).
Let us also assume for simplicity 
that the central source does not rotate. Then, the wind from the 
central 
nonrotating source is surrounded by the wind from the accretion disk which 
is in Keplerian rotation. According to the asymptotic theorem, in the wind  
from the disk there should 
exist at least one streamline which is collimated exactly along the axis of 
rotation. But 
then, automatically all the mass flux enclosed by this streamline will 
be collimated too.
It is important to note that in this case there is no
physical limitation on the amount of collimated mass flux.
All mass from the central nonrotating part of the outflow is
expected to be collimated in a jet. 

The slow rotation of the central part of the source of the wind is 
apparently not the only way to increase the amount of the mass flux in
the jet.  
Let us have a look on the problem from another point of view. 
The collimation of the plasma is controlled by the Lorentz force which acts 
accross the field lines. Using a local coordinate system with unit vectors 
$(\hat p, \hat \varphi, \hat n)$ where $\hat p$ is tangent to the poloidal 
field line and $\hat n$ is directed towards the local centre of curvature 
of the poloidal line, the Lorentz force along $\hat n$ is,
\begin{equation}
F_n= \frac{J_pB_{\varphi}-J_{\varphi}B_p}{c} 
\,.
\label{lf1}
\end{equation} 
{The first term in this expression represents the collimating  
force  produced by
the toroidal magnetic field $B_{\varphi}$. The second  term  corresponds
to the decollimating force from the poloidal magnetic field.
The collimating Lorentz force is as follows
\begin{equation}
F_n=\frac{1}{4r} \frac{B_p\partial (rB_{\varphi})^2}{\partial\psi}.
\label{lf}
\end{equation} 
where $d\psi = 2 \pi r B_pd l$, $r$ - is the distance  to  the  axis  of
rotation, $B_p$ is the poloidal magnetic field, $dl$ is the 
length element across the magnetic field. The toroidal magnetic field can be 
estimated from the frozen in condition 
as} 
\begin{equation}
B_{\varphi}\approx {r\Omega B_p\over v_p}.
\label{bf}
\end{equation} 
provided that we consider the flow far enough from the Alfvenic surface, 
and, therefore, 
we can neglect the toroidal velocity of the plasma in comparison to the 
poloidal velocity.
This is a good approximation for analytical solutions with radial poloidal 
streamlines (Tsinganos et al. 1992, Figs. 2,3)
and also in numerical simulations as in Krasnopolsky et al. (2000, Fig. 3) and 
Ustyugova et al. (2000, Fig. 10).
Let us introduce  the following definition
\begin{equation}
\mu=(r^2\Omega B_p/v_p)
\end{equation}
It follows from (\ref{lf})  that the derivative 
${\partial\mu^2/ \partial\psi}$ defines the collimating force. If 
${\partial\mu^2/ \partial\psi} > 0$  the Lorentz force is directed
towards the axis of rotation. The plasma in these regions is collimated. 
Otherwise the plasma
is decollimated. In general, both such regions can exist in magnetized 
winds (see for example Fig. 10 in Paper I).
A similar effect of partial decollimation of the wind has been found 
in Beskin \& Okamoto (2000) and also  Ustyugova et al. (2000). 
{It follows from
the previous asymptotic theorem that if $\mu \approx 0$ in the central part of 
the flow while it increases outwards, then all 
this part of the flow will  be  collimated.   A strong  decrease  of  this
parameter in the central part of the outflow can be achieved also if
the poloidal magnetic field is close to zero in 
the central
part of the flow or the velocity is drastically increased. It seems 
reasonable to assume that the efficiency of the collimation would be
increased in these cases as well.

All these assumptions need to be carefully verified.} 
In this paper we verify only the prediction of the asymptotic theorem 
that arbitrary
large flux of plasma and magnetic field can be collimated 
into the jet provided that 
the angular velocity in the central part of the flow is negligible.

\subsection{Model}

To verify the above conclusion that with a zero rotational velocity  
in the central part of the flow the mass loss in the jet
can be remarkably increased if it is surrounded by a rotating 
envelope, we use the simplest model wherein we have a 
cold wind outflow from a spherical source.
The initial
magnetic field is taken to be that of a split-monopole. The details of this 
model are described in Paper I. The justification  
why such a model can be applied for investigating the general properties 
of magnetic collimation in a wide class of
outflows is 
given in Paper II. In Papers I, II, the solution
of the problem of the selfconsistent steady state plasma outflow 
was divided in two stages. At the first stage the problem was solved
in the nearest zone by numerically simulating the  
time dependent plasma outflow. When the initial flow relaxed to a 
final steady state outflow, this solution was used for obtaining
a steady state outflow in the far zone.

\subsection{Results}

According to our previous qualitative analysis, it is expected that the 
fraction of the mass collimated into the jet will be increased in comparison 
to the case of a constant or accretion disk-like rotation,  if 
the angular frequency function $\Omega(\psi)$ is negligible in the central 
parts of the flow.
To verify this conclusion, we calculated the steady state of an outflow 
from a differentially rotating object wherein the angular velocity increases 
with the magnetic flux $\psi$. 
In Fig. 3 the poloidal lines of the flow are plotted in the nearest 
zone of the central object 
which rotates according to the following law
\begin{figure} 
\centerline{\psfig{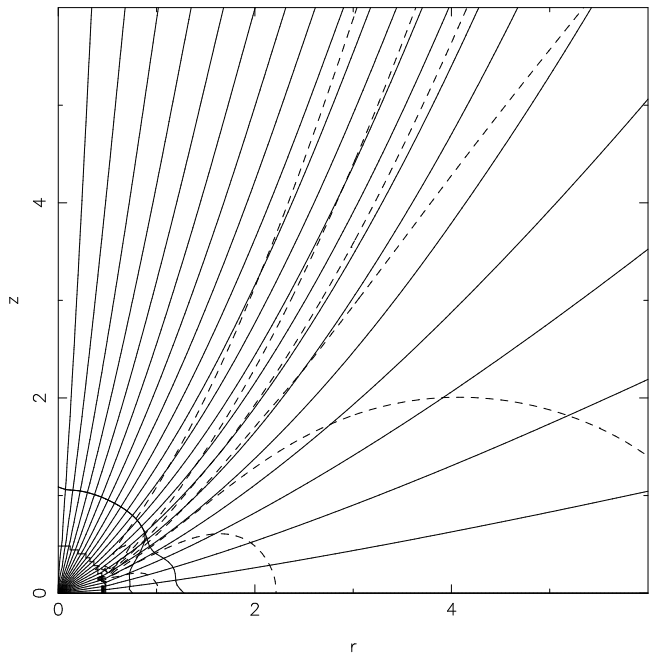}}
\caption{Poloidal lines of isotropic at the base wind from a  
differentially rotating object, wherein rotation is basically concentrated at
the equatorial region. The star is located at the lower left corner of the
figure. Solid lines denote poloidal magnetic field lines placed at equal 
angular distance from  each other at the isotropic outflow and dashed lines  
poloidal electric currents generating the toroidal magnetic field. Thick solid 
lines surrounding the star indicate Alfvenic and fast mode surfaces and split 
close to equator because basically only there a toroidal magnetic field is generated.
}
\end{figure}
\begin{equation}
\Omega (\psi) 
= \left\{ \begin{array}
{c@{\quad:\quad}l} 
4.5\times \displaystyle{\xi\over 7[50(\xi-0.7)^2+0.1]}{v_0\over R_f} & \xi < 0.7, 
\\  
\\ 
4.5 \displaystyle{v_0\over R_f} & \xi \ge 0.7, 
\end{array}\right.
\end{equation}  
where $\xi=\psi/\psi_{\rm max}$. According
to this law basically only the part of the source near the equator rotates 
while near the 
axis there is practically no rotation. The thick lines show the location of 
the Alfv\'en and fast mode critical 
surfaces. They are splitted only at the equatorial region since only in 
this region
a toroidal magnetic field is generated. It follows from this figure that 
as the 
rotation of the outer part of the central source is fast enough, the 
collimation of the 
flow from the central part occurs already in the nearest zone of the 
flow. 

\begin{figure} 
\centerline{\psfig{file=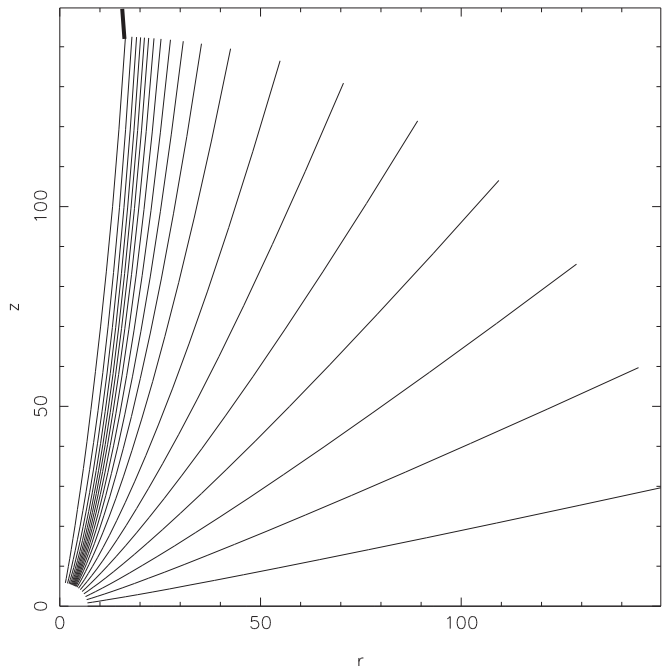,width=9.0truecm,angle=270}}
\caption{Poloidal lines of magnetic field of a differentially rotating 
wind as in previous figure but in the far zone.
In contrast to previous figure, the magnetic flux
enclosed by each field line differs by a constant value from line
to line. Thick solid line marks the beginning and assumed position
of a shock wave.
}
\end{figure}

\begin{figure} 
\centerline{\psfig{file=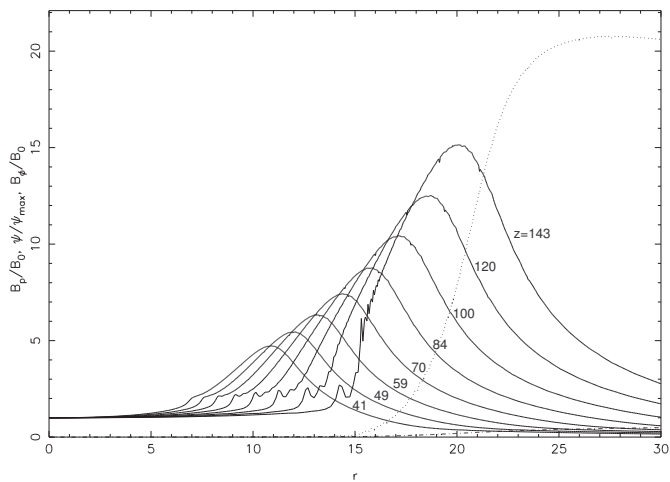,width=9.0truecm,angle=270}}
\caption{Distribution of magnetic field across the flow in a 
plane located at various labeled distances z across the rotation axis. 
The magnetic field is normalized to its value at axis. 
Sequence of curves shows the formation of a  
harp front in the distribution of the magnetic field which finally leads 
to the formation of a steady state shock wave in the flow at $z > 143$.
Dotted line denotes normalized toroidal magnetic field for the
largest z and dashed-dotted line the relative mass flux for the same
distance z. The jet contains more than $50\%$ of the initial total mass flux.}
\end{figure}

Fig. 4 shows the flow in the far zone. 
In this zone the collimation continues in a logarithmic scale as it was 
discussed
in Paper I. In addition however,  now we have the formation of a shock
wave in the collimated flow.  

The internal parts of the wind are not collimated by the  Lorentz force 
since
this layer of the outflow originates in a practically nonrotating central source. 
However, this wind is indirectly collimated by the outer layer of the outflow. 
This part of the wind affects 
the supersonic flow coming from the central part similarly to a curved wall 
in the flow of a supersonic gas (Landau \& 
Lifshitz 1975, p. 429). The shock is formed where the 
characteristics intersect. The beginning of the shock and  
its position 
in the flow are shown schematically in Fig. 4 by a thick solid line. 
The process of the 
formation of the shock front can be better seen in  Fig. 5, where 
the distribution of the normalized poloidal magnetic field {\it vs.} distance to 
the axis of
rotation is shown for different heights z from the central source. 
The poloidal 
magnetic field is raked by the toroidal magnetic field generated 
in the outer parts of the wind. But the central parts near the axis 
still do not
have received the signal about this raking. Therefore a wave-like 
structure of the 
poloidal magnetic field  is formed, as in Fig. 5. The numbers near the curves show the 
distance z expressed in units of the initial Alfven radius.
It is clearly seen that the shape of this wave-like structure of the 
poloidal magnetic field line changes with z. 
The front of the wave becomes more sharp with increasing z.
Finally, the front becomes practically vertical forming a shock. 
Unfortunately, the flow containing this shock cannot be calculated with 
our code, since we solve 
the steady state Cauchy problem in the two dimensional space of r,z (Paper I).

For the last z and just before the shock is formed, we also plot in this 
figure the 
dependence of the normalized toroidal magnetic field and the normalized 
poloidal
magnetic field flux. It can be seen from this figure that at the maximum 
of the toroidal 
magnetic field which
corresponds to the boundary of the jet which will be finally formed 
after the 
passage of the shock front from the flow, the collimated flux 
exceeds $50\%$ of the total initial flux. For comparison, 
for jets from sources with uniform or disk-like rotation laws (Paper I), 
we had at maximum about $1\%$ of the flux collimated.  

The formation of the shock front in  the collimated wind 
occurs only if the characteristics of one 
family begin to cross at some point (Landau \& Lifshitz 1975). 
But this crossing does not happen in every flow. If the central part of the 
flow is 
compressed by the outer part of the 
wind close enough to the central source the shock wave is not formed. 
For example,
with th efollowing form of the angular velocity distribution, 
\begin{equation}
\Omega = 3.5 \xi^2 {v_0 \over R_f}
\,,
\label{22}
\end{equation}
the shock wave 
is not formed. The flow from this source in the far zone is shown in  
Fig 6. The jet from this source indeed is similar to observed astrophysical 
jets.
The corresponding to Fig. 5 distribution of physical quantities 
with distance from the jet axis in this case is shown in Fig. 7. 
The jet has a structure 
which 
remarkably differs from  the structure of the jets from the 
uniformly or disk-like rotating objects discussed
in Paper I (see Fig. 1). 
Thus, the jet under consideration  has uniform poloidal magnetic 
field and 
in this specific case contains also a remarkable part ( $\sim 25\%$ )
of the total magnetic and mass flux.

\begin{figure} 
\centerline{\psfig{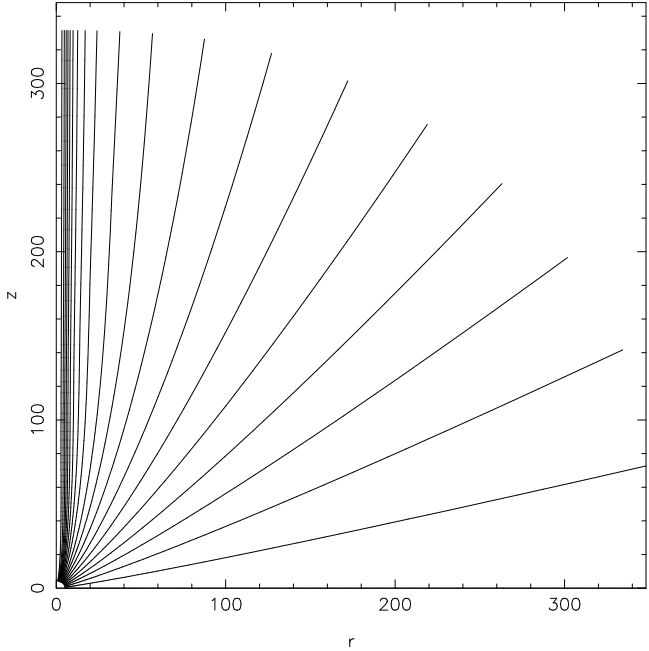}}
\caption{Poloidal lines of the flow in the far zone of the wind from an 
object rotating with the law given by Eq. (\protect\ref{22}).  The  flux
difference between the field lines is constant. A shock
wave is not formed in this case.}
\end{figure}
\begin{figure} 
\centerline{\psfig{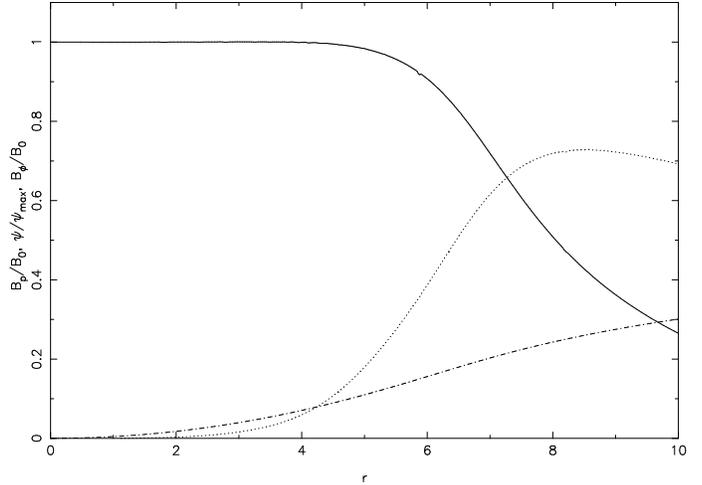}}
\caption{Distribution of the normalized poloidal (solid line) and
toroidal magnetic field in the jet formed by the differentially rotating 
source, as in Eq. (\protect\ref{22}). The jet formed in this
model has a rather uniform inner structure.}
\end{figure}

\section{Summary and Discussion}

According to recent studies on the spontaneous magnetic collimation of MHD winds, 
the flow of plasma at large distances from the source consists of a jet 
and a radially expanding wind (Paper I). The density of the
plasma in the wind drops down with distance while the jet is mixed 
with the surrounding circumstellar medium over a distance
small compared to its length. Since the density of the plasma
in the distance of the terminating shock is rather small, 
this process is hard to be observed directly. Therefore for an estimate
of the mass flux ratio $\dot M_{\rm jet}/\dot M_{\rm wind}$ we use indirect methods 
estimating  
the ratio  $\dot M_{\rm jet}/\dot M_{\rm accr}$. A comparison of the observable 
characteristics of jets from 
some YSO's provides evidence that the mass loss rate in the jets 
of these objects is a remarkable fraction of their total mass loss; 
this observed high mass loss ratio in jets greatly exceeds the prediction 
of the  simplest models. However, these observational
data can be explained if we assume that the characteristics of the 
source of the wind in the central part of the outflow 
differ remarkably from the characteristics of the surrounding wind. 
In principle, an arbitrary large fraction of the wind can be indirectly 
collimated into a jet
when the central part of the source does not rotate significantly and it is 
surrounded by a rapidly rotating envelope. Our numerical simulations totally 
confirm this conclusion, leading in some cases to a collimation of up to 
$50\%$ of the flux of the wind into the jet. 
 
The idea that a jet with a relative high mass loss fraction should be 
launched from the central parts of a source which produces an outflow
with properties strongly different from those of the surrounding wind, actually
is not new. Several studies have led to the same conclusion, albeit from 
different perspectives. For instance, the extraction of the rotational 
energy of a central black hole by a magnetic field and its eventual conversion  
to Poynting flux and relativistic $e^+-e^-$ pairs in the Blandfrod \& Znajek 
(1977) mechanism, is now a rather key element in models of AGN's 
(Sol et al. 1989, Pelletier et al 1996). Also, 
the idea that the observed jets are the result of the interaction of a 
central star with an accretion disk was proposed in the context of young 
stellar objects by Shu et al. (1991). 
 
What conditions can actually be realized in the central parts of a 
wind to produce jets with a high mass outflow?
Our analysis has shown that the slow rotation of the central 
part of the source of the wind is not the only way to increase 
the fraction of the mass lost in the jet. Apparently the same effect 
can be obtained by decreasing the poloidal magnetic field in the 
central part, or by strongly increasing the plasma velocity.
These are possibilities which should be investigated in more detail.

First, it is unlikely that the magnetic field in the central parts of the 
source is decreased remarkably. Most models on plasma ejection 
from accretion disks predict the opposite tendency : the poloidal magnetic 
field is rather enhanced at the central parts of the source (Oyued \& Pudritz 1997, 
Ustyugova et al. 2000, Krasnopolsky et al. 2000).
Therefore we are left with the following possibilities:
either the central angular velocity of rotation is decreased, or, the 
velocity of the plasma in the central parts of the source is increased.  
Of course a combination of these two conditions is also possible. 

The idea that the source of the jet rotates rather slowly may be quite
reasonable, at least in relation to YSO's. It is evident that
a protostar should rotate more slowly than the inner edges of its 
Keplerian accretion disk and observations indeed confirm this prediction.
We do not intend to argue here that the matter in the jet is ejected from 
the protostar. The close disk-jet connection shows that the matter in 
the jet is supplied by the accretion disk (Livio 1999).
But it is reasonable
to assume that this matter penetrates in the magnetic field of the 
central star, partially falls down on the surface of the star and 
partially is ejected outwards (Shu et al. 1991). 
In this case only the magnetic field 
of the jet is connected with the central star. Schematically this picture 
of the outflow is presented in Fig. 2. According to this scheme the disk
not only supplies the plasma of the jet, but also it produces the magnetized 
wind which collimates the outflow from the central source into a jet. 

Another possible configuration for the formation of jets with relative 
high mass outflow is connected to the possibility of an outflow from the  
central  source with speeds much higher than those of the surrounding wind (see 
also Paper II). 
It is important to note that this mechanism  allows naturally 
the collimation of a relativistic plasma. In Paper I it was shown that
a relativistic plasma is practically not collimated by the magnetic 
 
stresses even at very fast rotation (see also, Bogovalov 1997, Bogovalov 2000).
This is a rather intrinsic property of relativistic outflows.

There are two reasons why a relativistic plasma cannot be selfcollimated.
In a relativistic outflow the role of the electric field becomes important.
According to the frozen in condition, the electric field is ${\bf E} =
-{\bf v \times B}/c$. In the nonrelativistic limit this electric
field is negligibly small and therefore the Coulomb force  
on the induced in the plasma electric charge is also negligible. 
In the relativistic limit the situation changes. The Coulomb force becomes
comparable to the Lorentz force but it affects the plasma  in the 
opposite direction (Bogovalov 2000), drastically decreasing the
effect of
the magnetic collimation. The collimating force increases
with the angular velocity of the rotation of the central source. 
However, even at very fast rotation rates the collimation remains very inefficient 
since another relativistic  effect comes into play. The effect of 
collimation depends not only on the force pushing the plasma to 
the axis of rotation, but also on the mass of the plasma. In the 
relativistic limit the flux of the energy of the electromagnetic field
at fast rotation exceeds the flux of the energy of the matter. But 
according to the relativistic relationship between energy and mass,
the flux of the energy is equivalent to the flux of matter.  
The effective mass of the outflowing relativistic plasma is effectively
increased by the contribution of the mass of the electromagnetic field.
Therefore  the increase of the angular velocity of the rotation or 
the magnetic field does not provide a more efficient collimation of the 
relativistic plasma. The effective mass of the plasma increases almost
proportionally to the collimating force. 

In the proposed scenario, which may be applicable for example to blazars, 
the relativistic plasma is ejected   
from a central source, for example a black hole, e.g., via the Blandford \& 
Znajek (1977) mechanism. Then, the collimation of this
plasma is provided by the magnetic nonrelativistic wind from the 
accretion disk, as shown in Fig. 2.  A more detailed verification of this
mechanism for the collimation of relativistic plasmas will be presented in another 
connection.

\medskip\noindent
{\bf Acknowledgements}. The authors are grateful to Dr S. Lamzin for 
a duscussion of observed properties of 
outflows from  YSO's and an anonymous referee. This work was supported in part by grants 
NATO grant CRG.CRGP 972857,  INTAS-ESA N 99-120   
and the Ministry of Education of Russia 
in the framwork of the program
"Universities of Russia - basic research", project N 990479.

\end{document}